\documentstyle[12pt,aaspp4,psfig]{article}
\begin{document}
\title{Cataclysmic Variables from SDSS I. The First Results \footnote{Based on 
observations obtained with the Sloan Digital Sky Survey and with the
 Apache Point
Observatory (APO) 3.5m telescope, which are owned and operated by the
Astrophysical Research Consortium (ARC)}}

\author{Paula Szkody\altaffilmark{2}, Scott F. Anderson\altaffilmark{2}, 
Marcel Ag\"ueros\altaffilmark{2},  
Ricardo Covarrubias\altaffilmark{2},
Misty Bentz\altaffilmark{2}, Suzanne Hawley\altaffilmark{2}, 
Bruce Margon\altaffilmark{2,3}, Wolfgang Voges\altaffilmark{4}, 
Arne Henden\altaffilmark{5}, Gillian R. Knapp\altaffilmark{6},
Daniel E. Vanden Berk\altaffilmark{7}, Armin Rest\altaffilmark{2}, 
Gajus Miknaitis\altaffilmark{2}, E. Magnier\altaffilmark{2}, 
J. Brinkmann\altaffilmark{8}, I. Csabai\altaffilmark{9},
M. Harvanek\altaffilmark{8},
R. Hindsley\altaffilmark{10}, G. Hennessy\altaffilmark{10},
Z. Ivezic\altaffilmark{6}, S. J. Kleinman\altaffilmark{8},
D. Q. Lamb\altaffilmark{11},
D. Long\altaffilmark{8}, P. R. Newman\altaffilmark{8},
E. H. Neilsen\altaffilmark{7},
R. C. Nichol\altaffilmark{12}, A. Nitta\altaffilmark{8}, 
D. P. Schneider\altaffilmark{13}, S. A. Snedden\altaffilmark{8},
D. G. York\altaffilmark{11} }


\altaffiltext{2}{Department of Astronomy, University of Washington, Box 351580,
Seattle, WA 98195}
\altaffiltext{3}{Space Telescope Science Institute, Baltimore, MD 21218}
\altaffiltext{4}{Max-Planck-Institute f\"ur extraterrestrische Physik,
Geissenbachstr. 1, D-85741 Garching, Germany}
\altaffiltext{5}{US Naval Observatory, Flagstaff Station, P. O. Box 1149,
Flagstaff, AZ 86002-1149}
\altaffiltext{6}{Princeton University Observatory, Princeton, NJ 08544}
\altaffiltext{7}{Fermi National Accelerator Lab, P.O. Box 500, Batavia, Il 60510}
\altaffiltext{8}{Apache Point Observatory, P.O. Box 59, Sunspot, NM 88349-0059}
\altaffiltext{9}{Dept. of Physics and Astronomy, John Hopkins University,
3701 University Drive, Baltimore, MD 21218}
\altaffiltext{10}{US Naval Observatory, 3450 Massachusetts Ave, NW, Washington,
D.C. 20392-5420}
\altaffiltext{11}{Dept. of Astronomy and Astrophysics and Enrico Fermi
Institute, 5640 S. Ellis Ave, Chicago, IL 60637}
\altaffiltext{12}{Dept. of Physics, Carnegie Mellon University, 5000 Forbes Ave,
Pittsburgh, PA 15232}
\altaffiltext{13}{Dept. of Astronomy and Astrophysics, The Pennsylvania State
University, University Park, PA 16802} 

\begin{abstract}

The commissioning year of the Sloan Digital Sky Survey has demonstrated that
many cataclysmic variables have been missed in previous surveys with
brighter limits. We report the identification of 22 
cataclysmic variables, of which 19 are new discoveries and 3 are known
systems (SW UMa, BH Lyn and Vir4).  A compendium of
positions, colors and characteristics of these systems obtained from the
SDSS photometry and spectroscopy is presented along with data obtained 
during follow-up studies with the Apache
Point Observatory (APO) and Manastash Ridge Observatory (MRO) telescopes.
We have determined orbital periods for 3 of the new systems: two show
 dwarf nova outbursts, 
and the third is a likely magnetic system with eclipses of its 
region of line emission. Based on these results, we expect the completed survey
to locate at least 400 new CVs. Most of these will be faint systems with low
accretion rates that will provide new constraints on binary evolution models.  
\end{abstract}


\section{Introduction}

Cataclysmic Variables (CVs) comprise all the close binaries that contain a
white dwarf accreting material transferred from a companion, usually
a late main-sequence star (see review book by Warner 1995). CVs include
novae, dwarf novae and nova-likes, which are distinguished by their amplitudes
and timescale for variability. A nova outburst due to a thermonuclear runaway 
event creates a rise in brightness of 10-20 mag; a dwarf nova undergoes
 2-7 mag outbursts from a normal quiescent state on a quasi-periodic timescale
ranging from weeks to years (usually 
attributed to a disk instability); and a nova-like exhibits 
random variability of several magnitudes likely related to changes in mass
transfer causing states of low or
high accretion.  
A large range in
orbital period (from just under 80 min to over 2 days, with most being under
2 hrs) and in magnetic field strength of the white dwarf (from $<$ 1 MG to 
240 MG),
results in a wide variety of characteristic behavior and various 
 observational
selection effects. While most of the nearly 1000 known or suspected CVs
 (Downes et al. 2001; D01)
are faint at quiescence and have late-type M dwarf secondaries, 
most of the early
discoveries from past surveys were blue, bright objects with dominant accretion
disks or outbursts that brought them to high luminosities, thus making them
accessible to detection. 

Evolutionary models for close binaries (Howell, Rappaport \& Politano 1997) 
predict CVs 
should reach a period minimum near 80 min
in the lifetime of the Galaxy and have very cool, optically unobservable
secondaries with very low mass transfer rates. Surveys that have the
capability to reach faint magnitudes ($>$ 16) should be able to find this
population of faint, old systems.
By determining the numbers and characteristics of these faint systems, we can
constrain
evolutionary theories and determine the true distribution of CV types in the
Galaxy. We expect these results to be quite different from the bright-end
surveys comprised of accretion disk-dominated
systems e.g. the 
Palomar-Green survey (Green et al. 1982).

The Sloan Digital Sky Survey (SDSS, York et al. 2000) 
is ideal for finding these faint blue and
red objects. The photometry of 10$^{4}$ deg$^{2}$ of sky in 5 filters to
$>$20th
mag allows a selection of all types of CV
systems (blue ones dominated by disks or white dwarfs as well as those that are
both blue and red if the accretion disk is negligible and the primary white
dwarf and M dwarf secondary are the main sources of light). 
In addition to the photometry, SDSS spectroscopy allows
unambiguous identification of a CV from the strong hydrogen Balmer and helium 
emission lines
that typically signify ongoing accretion.  The strength and width of these
lines provide clues to the inclination, orbital period and mass accretion rate
(Patterson 1984, Warner 1995). 
We report here the results we have obtained for objects we could clearly
identify as CVs during the SDSS commissioning year (data obtained prior to
1 January 2001) and the implications for the total number of CVs that will be 
found
by the completion of the SDSS project.

\section{Finding CV Candidates}

The SDSS imaging data are obtained on a dedicated 2.5m telescope at Apache
Point Observatory with a mosaic CCD camera (Gunn et al. 1998)
which observes in 5 bands $u,g,r,i,z$ (Fukugita et al. 1996; see
Lupton, Gunn, \& Szalay 1999 for a discussion of the modified magnitude
scale). Until the final SDSS photometric system is defined, the preliminary
magnitudes are denoted by $u^*,g^*,r^*,i^*,z^*$, with the errors in the
calibration 
estimated to be about 5\% at 20th mag. The data are reduced by a
photometric pipeline (Lupton et al. 2001) which separates stars from
galaxies using their surface brightness distribution and computes several
types of magnitude, including the point source magnitudes which we use in this
paper. The astrometric accuracy is 0.10 arcsec. Spectra of objects selected from
the photometric data are observed for typically 45 min 
with two fiber-fed CCD spectrographs
(on the 2.5m telescope)
covering the range of 3900-6200\AA\ (blue beam) and 5800-9200\AA\ (red
beam) at a resolving power of $\sim$1800 (York et
al. 2000). The spectra are wavelength and flux-calibrated and atmospheric
absorption bands are corrected using sdF stars. The resulting spectra are 
then classified as stars, galaxies or quasars and redshifts are determined. 
 Of the 640 fibers used in each 3$^\circ$ field, 32 are used for sky
measurements, $\sim$10 for standards and the rest are primarily for galaxies
and quasars. Stellar projects typically target about 10 fibers per field 
with CVs having approximately
one of these fibers.  

During the commissioning year, a number of photometric selection criteria were
attempted in order to maximize the chance of finding  
new CVs.  We first tried the range of colors found by
Krisciunas, Margon \& Szkody (1998), who used filters close to those of SDSS
to observe known CVs
 of various types. The CVs easily separate from  
main-sequence stars in $u^{*}-g^{*}$, but overlap with quasars (Fan 1999) 
and white dwarfs. 
In addition,
the large color range possible for redder colors does not allow a unique
discriminator. Thus,
selecting by the blue
colors resulted in spectra that turned out to be primarily
 quasars and single white dwarfs. When we tried simultaneous blue and red
criteria ($u^*-g^*<0.45, g^*-r^*<0.7, r^*-i^*>0.30$ and $i^*-z^*>0.4$) to
select for CVs without much disk contribution, we were successful in
obtaining mostly non-interacting WD+M binaries with a few CVs.
These criteria remain for the project color selection;  
in addition to the CVs found, the 
WD+M binaries
are being used in follow-up studies to search for systems which are close
to starting mass transfer (i.e pre-CVs), and to compare magnetic activity
levels for single M stars to those in close binaries (Raymond, Szkody \& Hawley
2001).

The best success at finding the blue
CVs came from using the fact that QSOs are primary targets for spectroscopic
fibers (hence more are observed than stars) and some quasars overlap the colors
of CVs. 
Thus, we could search the spectra obtained of quasar candidates
for those that turned out to be non-quasars i.e. those having zero-redshift and
broad hydrogen Balmer and helium
emission lines. 
Several fibers for extreme blue objects obtained under
a serendipity classification also yielded blue CVs. 

Because the commissioning year involved changing color selections for all 
targets, this is not a uniform sample. Nor will the final sample of CVs be 
complete,
as there are
not sufficient fibers for stars to target all possible CVs. However,
since the QSO, serendipity and CV selection criteria will not undergo further 
large
changes, the results from this past year cover a representative sample of
what will be found and represent a lower limit to the actual number of CVs
that can be located by combining the SDSS database with follow-up 
spectroscopic and variability studies. 
Our results reported here include spectra obtained from fields that
cover an effective area of about 600 sq deg. The spectroscopic 
identification of 22 CVs
in this area means that we found about 0.04 per sq deg. Thus,  we can
expect about 400 new CVs in the entire survey
 from the spectra alone. Follow-up time-resolved
spectra can be used to determine the nature of these objects, as well
as to identify pre-CVs from the WD+M binary sample. Wide-field,
time-resolved photometry
on selected fields will be necessary to identify a complete sample of CVs using 
the SDSS colors combined with variability information. 

\section{Results}

The 22 objects that we identify as CVs are listed in
Table 1 along with the date of the SDSS spectrum, and the $g^*$ magnitude and 
colors
(with no correction for reddening). The naming convention is 
SDSSp Jhhmmss.ss+ddmmss.s, where the
coordinate equinox is J2000 and the ``p'' refers to the preliminary
astrometry. Throughout the rest of this paper, we will abbreviate the names
to SDSShhmm for convenience. Objects which have data publicly available through
the SDSS early data release (Stoughton et al. 2001) are marked. 
Finding charts from the Palomar Digitized
Sky Survey (DSS) will be available from the online CV catalog (D01).
The last column of Table 1 gives brief comments which are elaborated in
the sections below. 

Figure 1 shows the colors of the 22 objects in the SDSS color-color diagrams
along with the stellar locus. As noted above, the contributions of the
accretion disk and white dwarf generally result in a separation of the
CVs from the main sequence, especially in the bluest colors. In this sample, 
about 90\%
of the objects were found by QSO and serendipity selection criteria. Two
CVs (SDSS0155 and SDSS0813) satisfied the selection criteria of blue+red
colors given in Section 2.

Figure 2 shows the SDSS spectra for all
systems except for
the previously known ones, BH Lyn and SW UMa, which have published spectra 
available (Hoard \& Szkody 1997; Shafter, Szkody \& Thorstensen 1986),
 and SDSS0155  which was 
saturated in the fixed exposure time for its field. 
Table 2 lists the equivalent widths and fluxes of the prominent hydrogen Balmer
and helium lines. In this table, the object is identified
both by its abbreviated SDSS name and the plate-fiber number, where the
plate number refers to the catalog number of the SDSS plug plate used to
observe each field. Of the 20 objects in Table 2, 13 (65\%) have H$\beta$
equivalent widths $>$20\AA, and 10 (50\%) have values $>$50\AA. According
to the approximate 
relation between H$\beta$ EW and $\dot{M}$ given in Patterson (1984), 
these numbers indicate we are finding the low accretion rate systems
($\dot{M}$ of 10$^{-10}$ to 10$^{-12}$M$_{\odot}$ yr$^{-1}$) i.e. the objects
are faint not because they are observed at a large distance, but because they 
are
intrinsically faint due to low mass transfer rates.
 
For several systems, we conducted follow-up
photometric observations
using the UW 0.76m telescope in Eastern Washington at Manastash Ridge 
Observatory
(MRO) equipped with a 1024x1024 CCD, and spectroscopic observations
with the UW share of the 3.5m telescope in New
Mexico at Apache Point Observatory (APO) equipped with a Double
Imaging Spectrograph (DIS). The spectrograph was used in high resolution mode
(resolution about 3\AA) with a 1.5 arcsec slit covering the wavelength regions
of 4200-5000\AA\ in the blue and 6300-7300\AA\ in the red. Sporadic spectra
on different nights were obtained to look for changes in the spectral lines 
and intensities that
would identify a dwarf nova outburst or the high/low state of a nova-like 
object.
For the brightest systems, time-resolved spectra over several hours 
were used to determine the
orbital period and to measure radial velocities. The orbital period can 
also be an indicator of accretion rate, as the lowest rates generally exist in
disk systems at quiescence that have orbital periods under two hours (Patterson 1984, Warner 1995),
although low accretion rates can also occur in systems with longer orbital
period when mass transfer sporadically turns off.
 The dates and modes of
follow-up observations are listed in Table 3.

\subsection{Previously Known Systems}

There were 3 previously known CVs for which SDSS spectra were obtained, as their
colors matched the selection criteria of quasars or serendipity. Two
are relatively well-known systems with published photometry and
spectroscopy (SW UMa and BH Lyn) listed  
in the D01 catalog as well as the Ritter \& Kolb (1998) catalog.
The third object (listed as Vir4 in the D01 catalog) 
is a recently discovered
eclipsing dwarf nova with an orbital period of 1.75 hrs (Vanmunster, Velthuis
\& McCormick 2000). Two SDSS spectra of this object at quiescence were obtained
(April and May; Table 1) and three APO spectra in May (Table 3).
As there are no published spectra of this
system, we show its spectrum (SDSS1435)
in Figure 1. All the spectra of Vir4 reveal deep central absorption in
the Balmer emission lines, which increases up the Balmer series. This feature is
 typical of high inclination, eclipsing systems (Vogt 1981) and is postulated to
be caused by low velocity, cool outer disk material in the line of sight.

\subsection{High Inclination Systems}

Two other systems also reveal the deep central absorption that is likely an
indication of high inclination (SDSS1555 and SDSS0151). 
Confirmation of high inclination from an
eclipse awaits time-resolved photometry or spectroscopy with larger aperture
telescopes, or
observations during an outburst when the system is brighter. 

\subsection{Dwarf Novae}

As the SDSS photometry provides a separate epoch of observation from the
SDSS spectroscopy, and the Palomar DSS provides an additional observation,
 there are several instances where the brightness of the system can be
verified to be 
different, signifying either outbursts (in the case of dwarf novae) or high
vs low states for nova-like objects. To distinguish these cases requires spectra
at the high and low brightness states.

In the case of SDSS1637, the SDSS photometry reveals a bright source 
($g^*$=16.6)
while the spectra are consistent with a 20th mag object. In addition, the
SDSS photometry reveals two close (but separated) stars, the northern 
object being the
bright one, while the DSS shows it fainter than its southern neighbor. This
fact is consistent with the northern object being caught at an outburst 
during the SDSS
photometry. Several random followup spectra at APO (Table 3) and other 
quick-look images
 during the course of
3 months all found the system at
its
quiescent state so the outbursts are likely infrequent or of short duration. 
As the quiescent spectra
show none of the high excitation lines of a novalike system, this object is
very likely a dwarf nova.

In two instances (SDSS1730 and SDSS2303), the APO
spectra revealed objects undergoing a dwarf nova outburst, during which
 the Balmer
emission lines turned into absorption lines when the systems were several
magnitudes brighter than quiescence.
 On 2000 October 2, we obtained a total of 20 time-resolved spectra of 
SDSS1730 
at outburst and two spectra, three nights later, when the system was at
quiescence. Figure 3 shows typical outburst spectra and the
smoothed mean of the 2 quiescent spectra for the blue and red wavelength
regions. At outburst, the overall flux
increased by a factor of 20 in the blue and 4 in the red, with noticeable,
broad absorption features surrounding the H$\beta$ and H$\gamma$ emission
lines. Figure 4 shows a closeup of the H$\beta$ line throughout the orbit.
While inspection of the line suggests the possible presence of 
a narrow component passing from blue (phase 0.48) to red (phase
0.57), 
careful analysis of the
velocity of the 
peak component did not produce an unambiguous feature which could be
associated with the orbital period.

The presence of the underlying absorption features at H$\beta$ and H$\gamma$
in SDSS1730 made it
difficult to obtain radial velocities for these two lines, and a reliable
solution was obtained only for the H$\alpha$ data. We obtained values for
the central peak of the H$\alpha$ line by using IRAF's
\footnote{{IRAF (Image
 Reduction and Analysis
Facility) is distributed by the National Optical Astronomy Observatories, which 
are operated by AURA,
Inc., under cooperative agreement with the national Science Foundation.}}
 fitting routines in 
splot.
As the line shows a fair amount of structure at outburst, the centroid
``e'' 
method returned more consistent answers than did Gaussian fitting. An IDL
routine, CURVEFIT, was used to fit the radial velocities with a 
sinusoidal function of the form
$v = \gamma - K sin[2\pi(t-t_0)/P]$, where $\gamma$ (systemic velocity), 
$K$ (semi-amplitude), $P$ (orbital period) and $t_0$ (time of conjunction) 
were left as free parameters
(see Table 4 for the best fit parameters). Figure 5 shows the best fit to
the H$\alpha$ radial velocity curve. Four spectra showed extreme
deviations and were not included in the fitting, and are indicated in
Figure 5 as open circles. We studied these spectra for evidence of
shifts in the lines and for other potential sources of error, but were
unable to detect any cause. Nevertheless, they provided velocities which were
clearly inconsistent with the rest of our measurements. The velocities for
these 4 spectra shown in Figure 5 are from measurements of the midpoint of
the extremes of the H$\alpha$ line done by eye. We tested our
fitting program while including these spectra, and found that the
best fit obtained for the orbital period (117$\pm$5 min)
 was consistent with the fit excluding these points. 

        Finally, we used the period obtained from the spectra 
to look for evidence of
variability in the 
 photometry of this system obtained at MRO.
We folded the differential
magnitudes on each of the two longest nights on the orbital period
 but found no evidence for any significant orbital 
variability. An average of the two nights' data produces an upper limit
for any such variability of 0.04 magnitudes. The absence of variability,
along with the lack of line doubling, suggests that SDSS1730 is a
system with a fairly low inclination (Warner 1995).

For SDSS2303, we obtained a total of 9 time-resolved spectra at
quiescence on 2000 October 2
and 11 spectra at outburst 3 nights later. Figure 6 
 shows typical
 outburst and quiescent spectra for both nights. At outburst, the overall flux 
increases by a factor of 120 in the blue and 60 in the
 red with broad absorption features surrounding the
H$\beta$ and H$\gamma$ emission lines. 

We obtained the velocities for the H$\alpha$ and H$\beta$ emission lines 
by using the
 IRAF fitting routines, with the Gaussian fitting routine ``k'' providing
 the best fit to a velocity curve (lowest residuals).
Once again, the CURVEFIT routine was used to fit the radial velocity
data with a sinusoidal function.
The lines of H$\alpha$ and $H\beta$ from the quiescent data provided the
most consistent solutions with the lowest errors. These solutions are given
in Table 4 and the fit to the velocity curve for H$\alpha$ is shown in
Figure 7. 
The outburst data resulted in  
consistent period and velocity curves. Using the best fit period of 100 min, 
the phases for H$\beta$ were computed with the time of the 
 red-to-blue crossing of the emission lines for H$\alpha$ as phase zero. 
The absence of any eclipse during quiscence or outburst 
 suggests that SDSS2303 is a system with a fairly low inclination.

To obtain further information about the disk at quiescence,
we computed a doppler tomogram.  Doppler tomography 
(Horne 1991) is a technique that allows imaging of the line-forming regions in a
binary system by combining the velocity profile information obtained at all 
binary
phases. To construct our tomogram, we used the Fourier-filtered back-projection
program provided by Keith Horne and modified for our computers and plotting 
needs by Donald Hoard. As we have no fiducial phasing from an eclipse or lines 
from 
the secondary, we used the phasing from the red to blue crossing of the
H$\alpha$ line solution (Table 4) at quiescence. This assumes that
 the emission lines
originate near the white dwarf and that zero phase is inferior conjunction of 
the secondary. Figure 8 shows the tomograms for H$\alpha$ and H$\beta$ with a
mass ratio of $q=0.25$ used to compute the Roche lobe and stream. While a
phase resolution of 0.1 is not suitable to determine any detailed disk 
structure, it can provide information on the disk extent and the stream.
The tomograms of SDSS2303 are consistent with a low mass-transfer rate
system with minimal disk emission and no evidence for a luminous hot spot
near the stream location. There is some evidence for emission regions near
the locations of the secondary (V$_{y}$=+150, V$_{x}$=0) and of the white
dwarf (V$_{y}$=-150, V$_{x}$=0), but better data are required to confirm this.

\subsection{Nova-likes with Strong HeII}

The 3 systems SDSS0155, SDSS0729 and SDSS0747 
reveal strong emission in HeII $\lambda$4686. While  SDSS0729 is very faint and has
very narrow lines, indicative of a low mass-transfer rate (possibly a magnetic)
system, SDSS0747 shows the bright blue continuum and weak Balmer
emission of an old nova with a high accretion rate. 

The system SDSS0155 caught our attention for further study as the SDSS
photometry and spectra indicated a very bright object, but the Palomar 
DSS showed
a much fainter star. The presence of strong emission at the bright state,
combined with the presence of HeII, suggests either an AM Her system with
a magnetic white dwarf or a high accretion rate
SW Sex star (characteristics of these types of CVs are reviewed by Warner 1995).

 Our followup spectroscopy at APO, one month after the SDSS spectrum,
showed the system also in its high state. Time-resolved
spectra throughout 3 hours showed asymmetries and narrow component structure
in the Balmer emission lines. Most important, an eclipse of the Balmer lines was
evident (Figure 9). When this occurred, the Balmer and HeI lines went into 
absorption while HeII remained visible but weaker,
 and the continuum did not appear to
be eclipsed. This is opposite to the behavior typically seen in an eclipsing
SW Sex system (e.g. BH Lyn, Hoard \& Szkody 1997) where the HeII line
(usually originating near the white dwarf) is deeply eclipsed along with
the continuum light, while the Balmer emission
lines (from the outer disk area) remain largely visible through the eclipse. 
Unfortunately, rising humidity levels necessitated a dome closure which
caused a 70 min gap in the data sequence. The datasets before and after the
dome closure cover intervals of 100 min and 60 min respectively.

Despite the gap, the velocities show a large and repetitive 
variation, allowing the determination of the orbital period to be 87 min.
This short period also argues against an interpretation as an SW Sex system
 since those
objects usually have orbital periods between 3-4 hrs.
The best fit to the radial velocity curve for HeII is shown in Figure 10 and
the solutions are given in Table 4 for that line as well as H$\beta$ and
H$\alpha$. The very large (400 km/s) semi-amplitude of the velocity curves
of SDSS0155 compared to the other disk systems in Table 4 is typical of
AM Her systems (Warner 1995) and strengthens the case for a magnetic 
interpretation for
SDSS0155.  

Using the convention of the red to blue crossing of the HeII line as phase zero,
the eclipse of the lines occurs at phase 0.7,
which is maximum redshift.
The most reasonable explanation for the lack of continuum eclipse is that
the eclipse is of a line-forming area in a magnetic CV system. Because the lines
are affected through more than one 10 minute exposure, the area must be
larger than just a spot on the white dwarf. The emission region may be
eclipsed by the secondary or 
self-eclipsed by the white dwarf.
To test the geometry further,
we constructed tomograms for the HeII and H$\beta$ lines, trying a phasing
based on a zero of 
the red to blue crossing of the HeII line (if its origin were on
the white dwarf) and, alternatively, considering zero phase to be the eclipse.
Figure 11 shows the results for the phasing of red-blue crossing, which produced
tomograms that are consistent with those of AM Her stars (Hoard 1999, Schwope
et al. 1999) with the brightest emission zones associated with the magnetic
settling region (lower left quadrant) and the ballistic stream as drawn for
a mass ratio of 1/3. Unfortunately, the usual prominent areas due to the white
dwarf and the irradiated secondary are not apparent and so the actual phasing
is ambiguous. Data obtained during a low state should allow the absolute
phasing to be determined by using the velocities of spectral lines from both the
secondary star and the white dwarf.   

\subsection{Systems Showing the Secondary Star}

Inspection of the spectra of Figure 2 shows that there are two systems which
have spectral signatures of the secondary. SDSS0813 shows the spectrum
of a K5-M0 star (Mg band near 5200\AA\ and TiO at 7100\AA), similar to the long 
period system TT Crt (Szkody et al. 1992). Thus, SDSS0813
is likely to also have a long orbital period. The other system, SDSS0816, also
shows evidence for the TiO band at 7100\AA. 
The appearance of these features 
indicates a relatively low contribution of the accretion disk and hence, a low 
mass-transfer rate system.

\subsection{ROSAT, 2MASS and LONEOS Correlations}

The newly identified CVs were cross-checked against the X-ray 
ROSAT All Sky Survey
(RASS; Voges et al. 1999, 2000), the infrared Two Micron All Sky Survey (2MASS),
and the optical Lowell Observatory Near Earth Object Survey (LONEOS; Rest et.
al. 2001). 

Besides the previously known
 bright system that is a 
known X-ray source (SW UMa; Shafter, Szkody \& Thorstensen 1986), 
5 of the new CVs
are detected in the RASS bandpass of 0.1-2.4 keV. Table 5 lists the X-ray 
count rates of these 5 
sources. The two objects identified as dwarf novae (SDSS1730 and SDSS2303) are
 both detected, as well as
the system with strong HeII that is likely to be magnetic (SDSS0155). As
about half the known dwarf novae are detected by the RASS (Verbunt et al. 1997;
V97),
it is not so surprising that SDSS1730 and SDSS2303 are detected.
However, it is unusual that the RASS count rates are identical for these two 
systems,
while their optical fluxes differ by 3 mags. Since V97 found that a higher
accretion rate corresponds to a lower X-ray to optical flux ratio, this would
argue that the fainter system SDSS2303 has the lower accretion rate (consistent
with its shorter orbital period) compared to SDSS1730, and that the lower
optical flux is not just due to a greater distance for SDSS2303 than SDSS1730. 
In addition, it is
remarkable that the two dwarf novae are brighter than the likely
magnetic system SDSS0155, as V97 and others have found that magnetic systems
 stand
out as the brightest soft X-ray sources among CVs. However, this can be
explained if SDSS0155 was in a low state of mass accretion at the time of
the RASS observations. We searched the existing X-ray data on the objects in 
Table 5 for any evidence of time variability and found a result
 only for SDSS0233.
In this case, the long pointing revealed variability of at least a factor of
4 on a time-scale of 500 sec. However, the low statistics and sampling do 
not allow any conclusion about periodicity.

Table 6 lists the 2MASS detections among the 19 new CVs (the known system
SW UMa is also detected). The bright dwarf nova SDSS1730 is among the
detections and shows IR colors typical of accretion disks (Szkody 1985). 
The system
showing a K-M dwarf secondary in the SDSS spectrum (SDSS0813) is also detected 
and has
 red colors consistent with a K star. 
SDSS1712 appears to have red colors (also evident in the
optical; Table 1) but does not show any spectral features from a late 
secondary (Figure 2). Thus, this faint object may be at a large distance
and/or may have significant reddening. 

While the primary targets of the LONEOS project are moving objects, the
existing database of a large fraction of the sky imaged in multiple epochs,
and calibrated to USNO-A2.0 red magnitudes, 
provides a good resource for time-variable objects such as CVs. We searched
the LONEOS database for our new SDSS CVs and found observations for the
objects listed in Table 7 (data on the known system BH Lyn also exist).
This table summarizes the range of variability evident along with the
number of measurements made and the number of nights on which observations
were obtained. Two of the four objects show evidence of large variability
(more than one magnitude). The dwarf nova SDSS2303 was near
quiescence on 2 of the 6 measurements and at brighter magnitudes on the
other 4 observations. Considering that 1 of our 2 APO 
measurements was also at outburst, 
the outburst timescale for this dwarf nova is likely to be very short.
The other object showing large variablility is the eclipsing, likely magnetic, 
system SDSS0155. This objects shows one mag variations among observations
obtained on a single night (likely due to the eclipse), as well as 3 mag
variations
in observations separated by months (likely due to high and low states of
mass transfer). The LONEOS database can ultimately be matched to the
SDSS photometric fields to help identify new CV candidates by their
variability.

\section{Conclusions}

The commissioning year of SDSS operation has shown that a wide variety of
interesting CVs may be easily identified. Included among these are a few 
bright systems missed in past surveys as well as members of the fainter
 population that
models predict. This year of partial operation has led to the discovery of
19 new CVs and the recovery of 3 known systems. While the objects fainter
than 20th mag at short orbital periods will need their orbital characteristics
determined by 8-10m class telescopes,  
the 15-19th mag ones can
be followed-up with 4m class or smaller apertures. Using sporadic and
time-resolved data, we have been able to identify two dwarf novae and one
interesting, high excitation, eclipsing system that is likely to be a magnetic
CV. The fact that the orbital periods of these 3 systems are all short
(under 2 hrs), and that more than half of all the CVs found have large
H$\beta$ equivalent widths, argue that we are finding lower accretion-rate
systems compared to previous surveys. 
These preliminary results show that the SDSS will be highly successful at
identifying new CVs of a variety of types, with at least 400 total expected
from the completed survey. Most 
of these should be faint, low accretion-rate systems, which will
provide 
insight into the nature and numbers of the true CV population in our
galaxy. 

\acknowledgments

The Sloan Digital Sky Survey (SDSS) is a joint project of The University of 
Chicago, Fermilab, the Institute for
Advanced Study, the Japan Participation Group, The Johns Hopkins University, 
the Max-Planck-Institute for
Astronomy (MPIA), the Max-Planck-Institute for Astrophysics (MPA), New Mexico 
State University, Princeton
University, the United States Naval Observatory, and the University of 
Washington. Apache Point Observatory,
site of the SDSS telescopes, is operated by the Astrophysical Research 
Consortium (ARC). 
Funding for the project has been provided by the Alfred P. Sloan Foundation, 
the SDSS member institutions, the
National Aeronautics and Space Administration, the National Science Foundation, the U.S. Department of Energy,
the Japanese Monbukagakusho, and the Max Planck Society. The SDSS Web site is 
http://www.sdss.org/. 

This publication makes use of data products from the Two Micron All Sky Survey,
 which is a joint project of the University
of Massachusetts and the Infrared Processing and Analysis Center/California 
Institute of Technology, funded by the
National Aeronautics and Space Administration and the National Science 
Foundation.

The authors are grateful to the Lowell Observatory Near-Earth-Object Search
for making their data available for the CVs on our list.

\begin{figure}
\psfig{figure=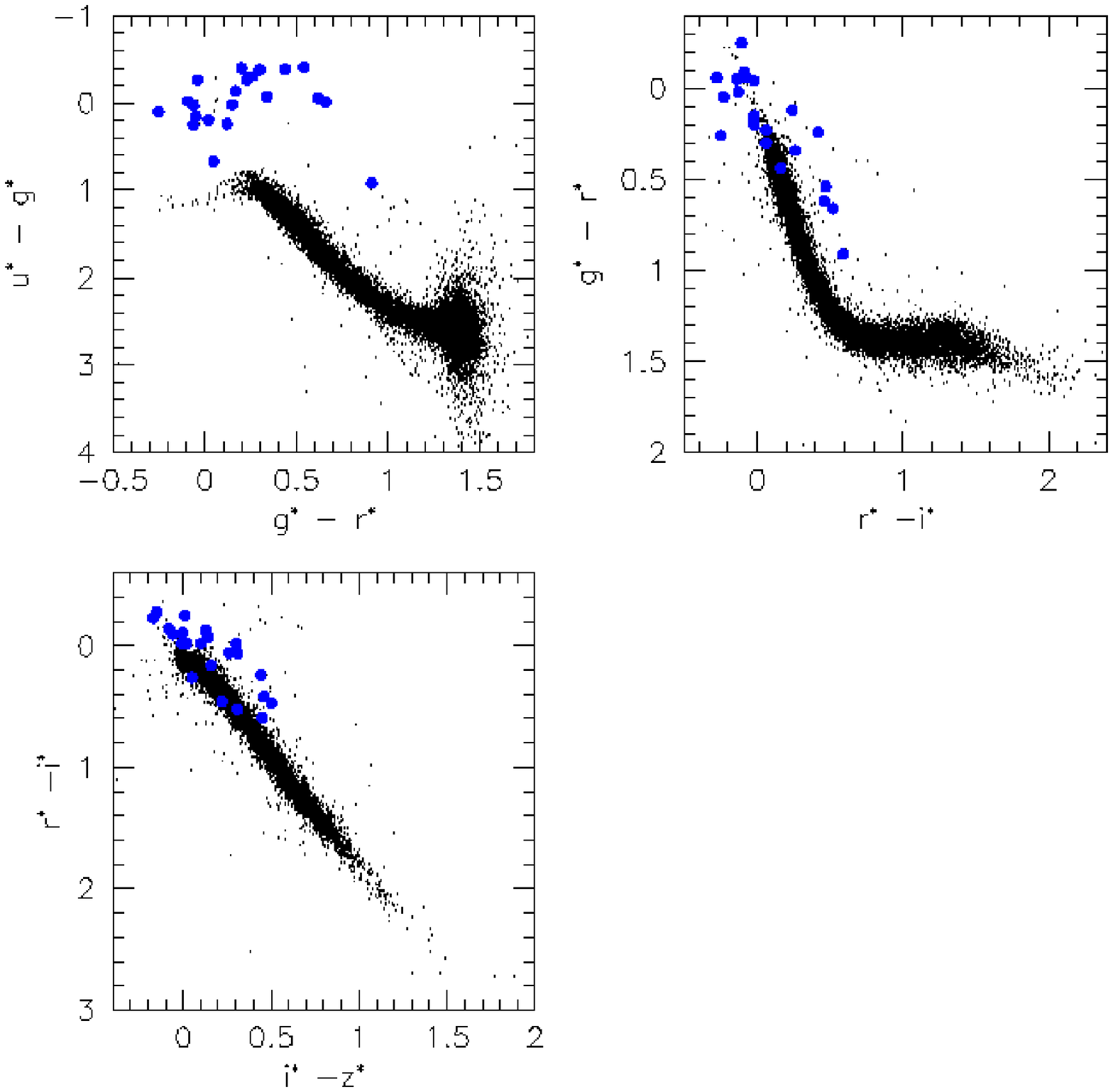,width=6in}
\caption{The SDSS color-color plots of the 22 objects in Table 1. The large
dots are the CVs while the small dots are stars defining the stellar locus.}
\end{figure}

\begin{figure}
\psfig{figure=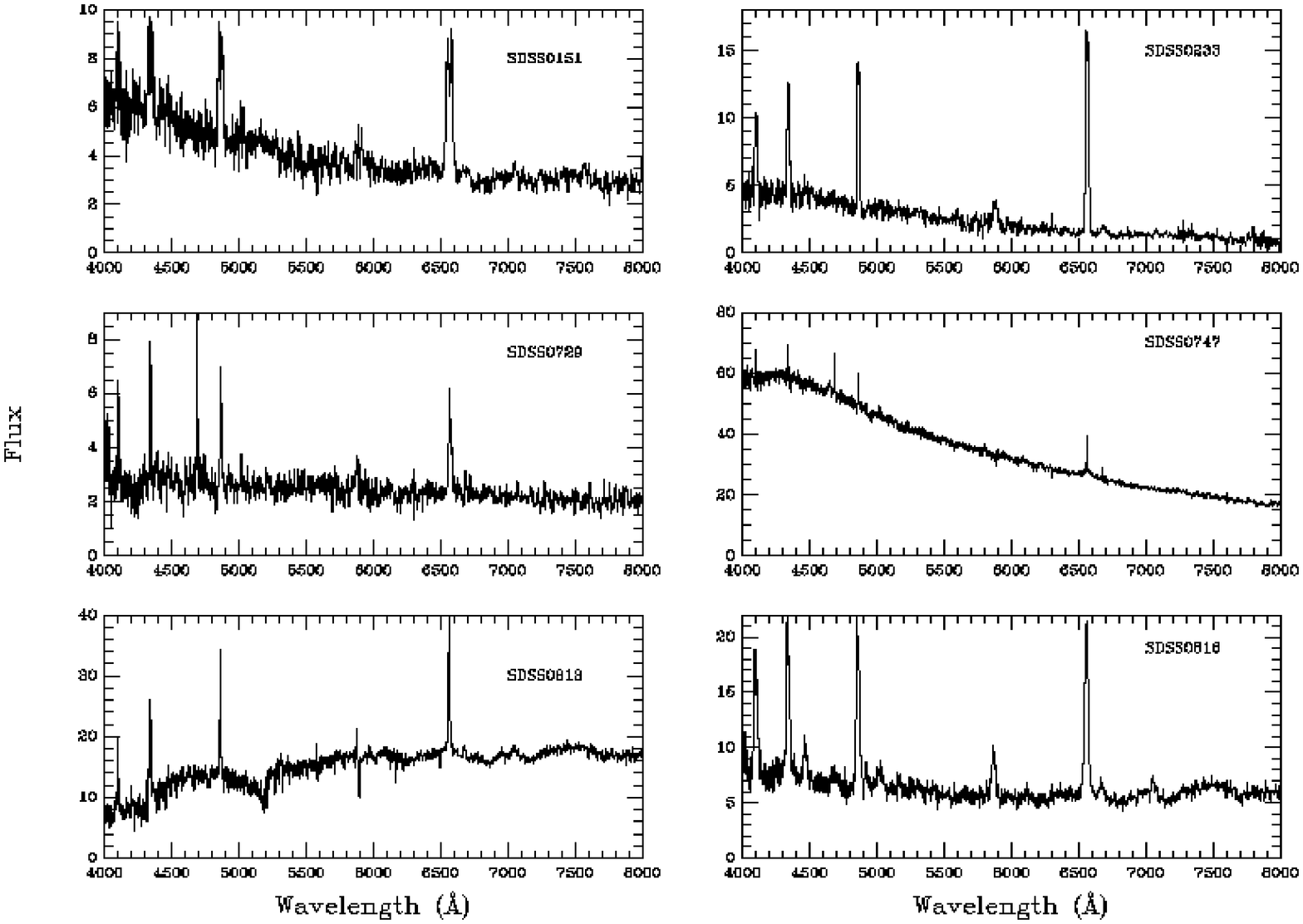,width=6in,angle=270}\psfig{figure=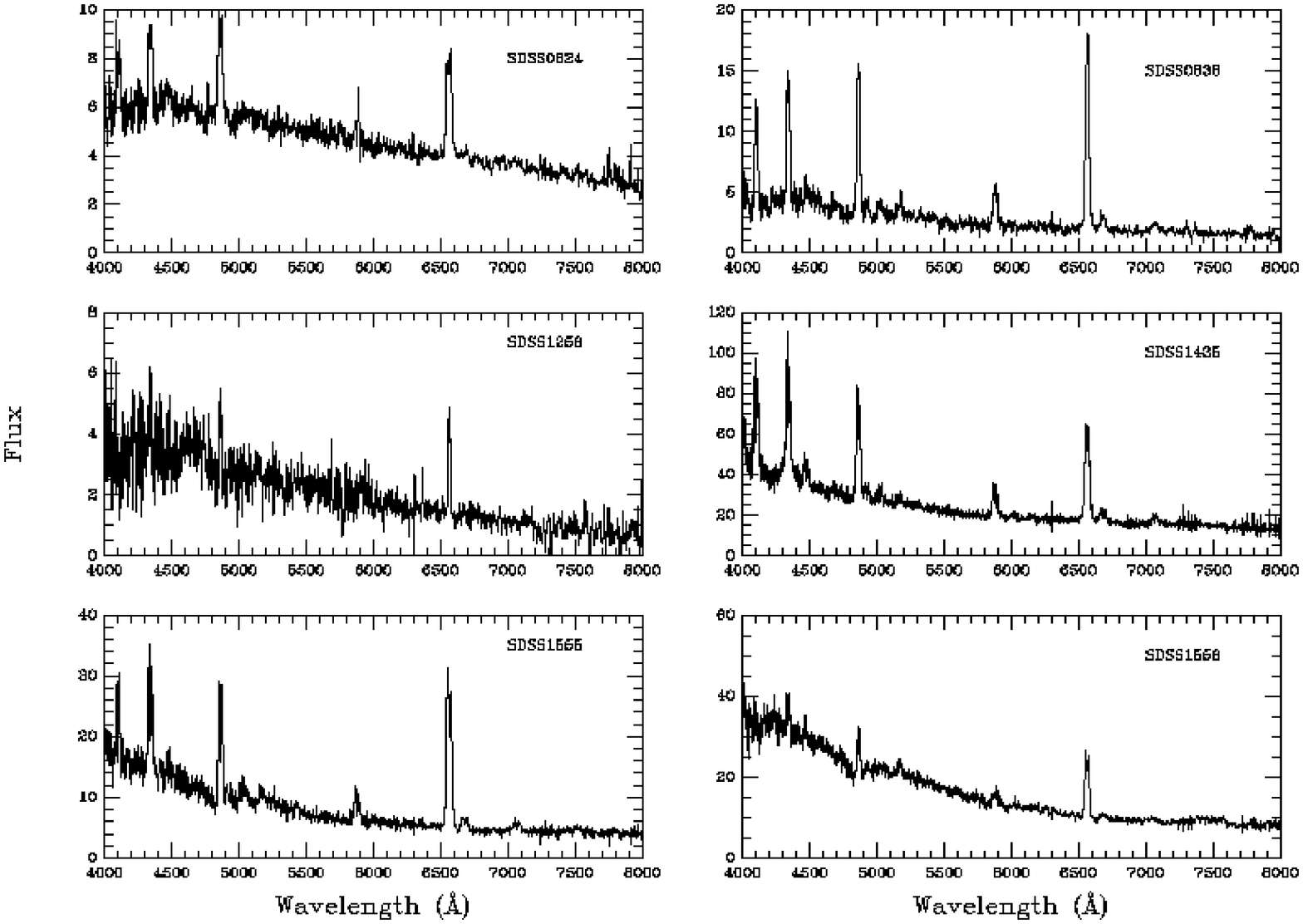,width=6in,angle=270}
\end{figure}

\begin{figure}
\psfig{figure=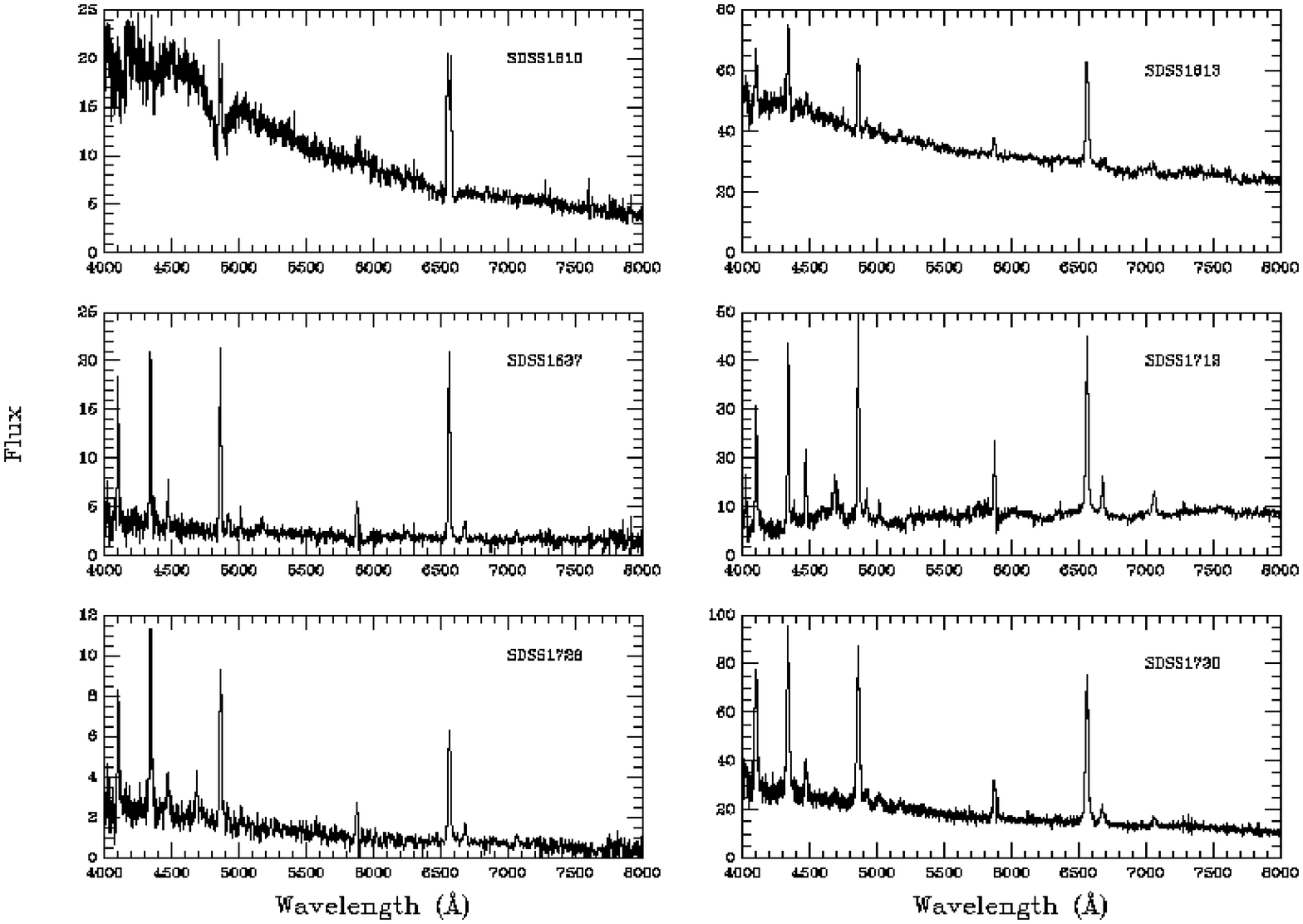,width=6in,angle=270}\psfig{figure=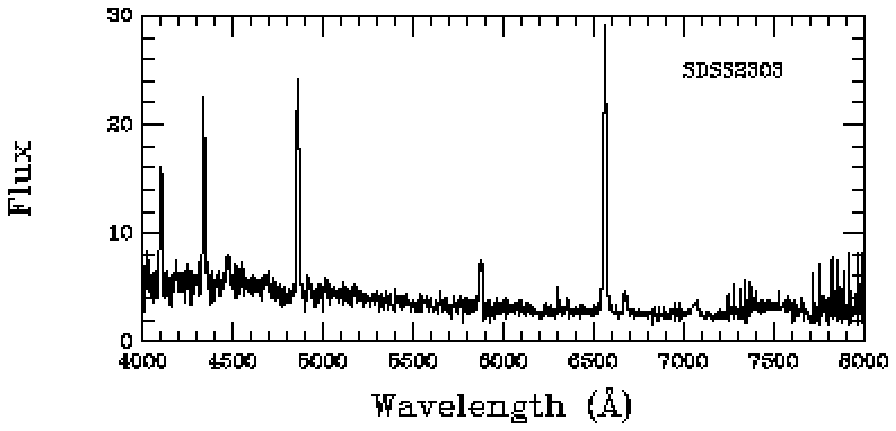,height=1.5in,width=3in,angle=270}
\caption{SDSS spectra of the newly discovered CVs. The flux scale is in units of flux density 10$^{-17}$ ergs cm$^{-2}$ s$^{-1}$ \AA$^{-1}$.}
\end{figure}

\begin{figure}
\centerline{\psfig{figure=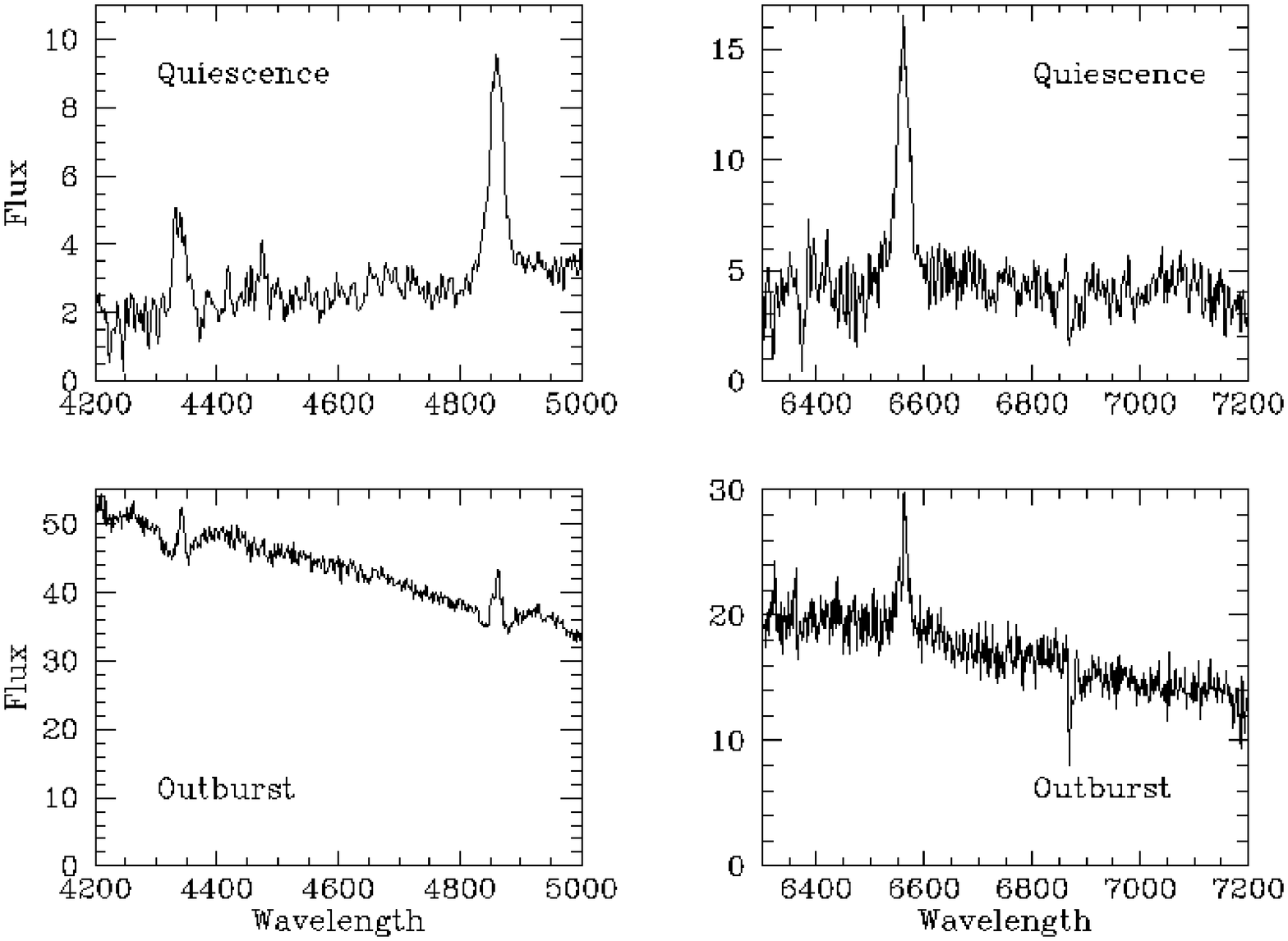,height=3.5in,angle=270}}
\caption{APO spectra showing the outburst and quiescence of SDSS1730. The fluxes have units of 10$^{-16}$ ergs cm$^{-2}$ s$^{-1}$ \AA$^{-1}$.}
\end{figure}

\begin{figure}
\centerline{\psfig{figure=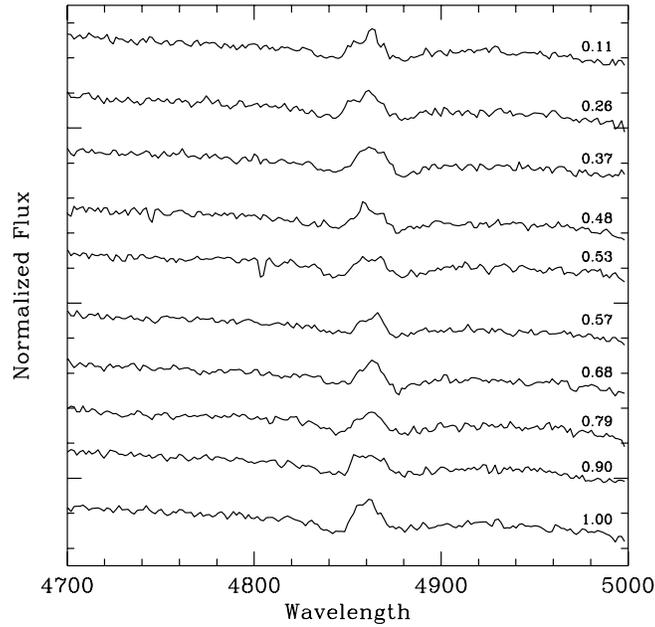,height=3.5in}} 
\caption{APO spectra showing the variation of the H$\beta$ line profile of SDSS1730. The fluxes have units of 10$^{-16}$ ergs cm$^{-2}$ s$^{-1}$ \AA$^{-1}$.}
\end{figure}

\begin{figure}
\centerline{\psfig{figure=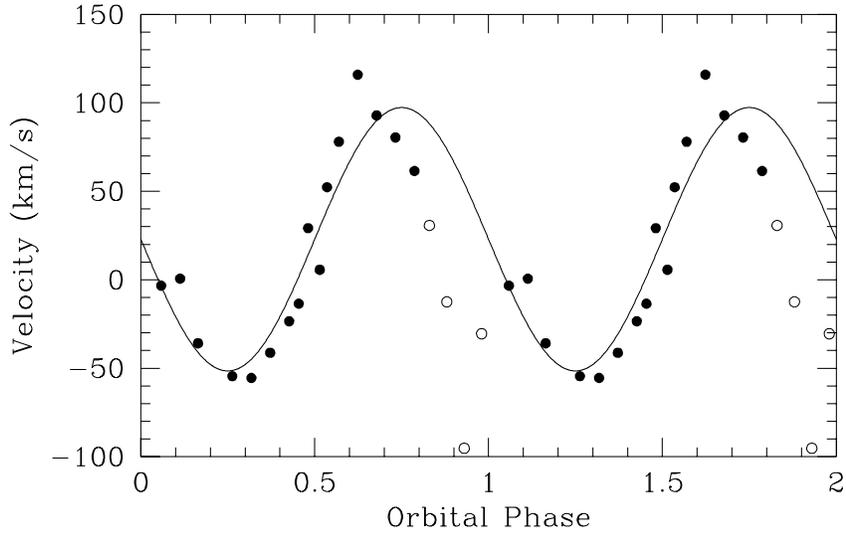,height=3in}}
\caption{The best fit to the radial velocity curve of SDSS1730 at
outburst using the emission line of H$\alpha$.
The open circles produced large deviations to a sine fit and were thus left
out of the best solution which is shown as the solid line.}
\end{figure}

\begin{figure}
\centerline{\psfig{figure=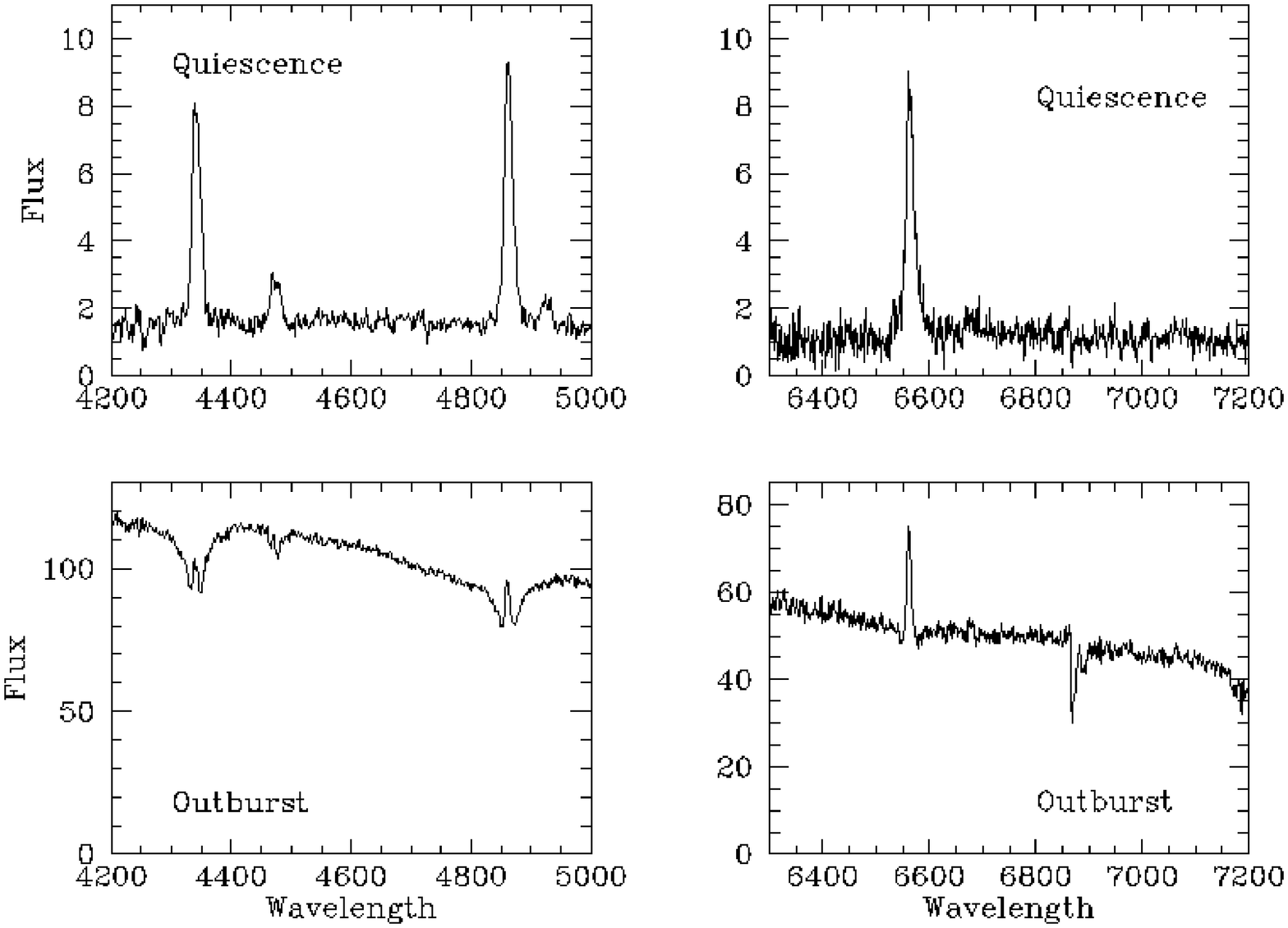,height=3.5in,angle=270}}
\caption{APO spectra showing the outburst and quiescence of SDSS2303. The
fluxes have units of 10$^{-16}$ ergs cm$^{-2}$ s$^{-1}$ \AA$^{-1}$.}
\end{figure}

\begin{figure}
\centerline{\psfig{figure=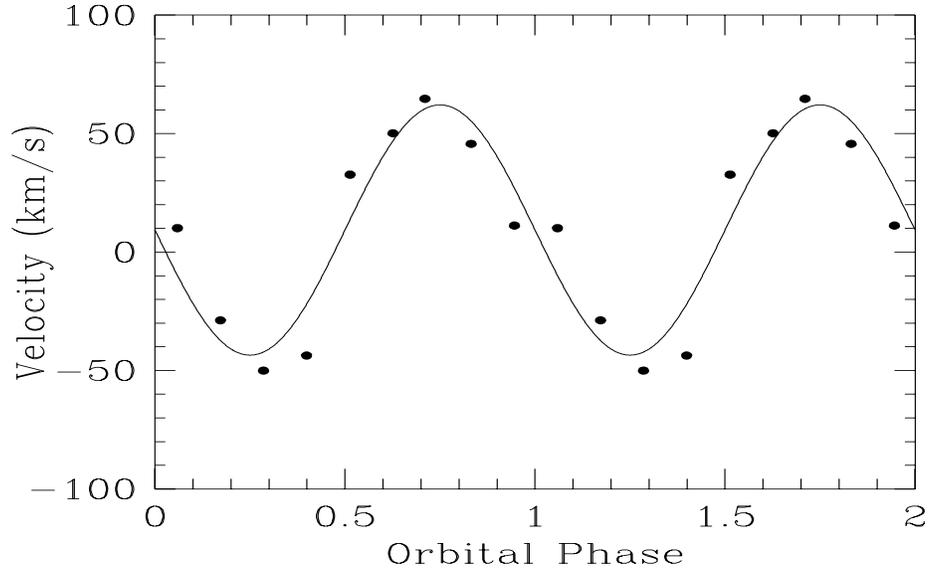,height=3.1in,width=5in}}
\caption{The best fit to the H$\alpha$ radial velocity curve of SDSS2303
at quiescence.}
\end{figure}

\begin{figure}
\centerline{\psfig{figure=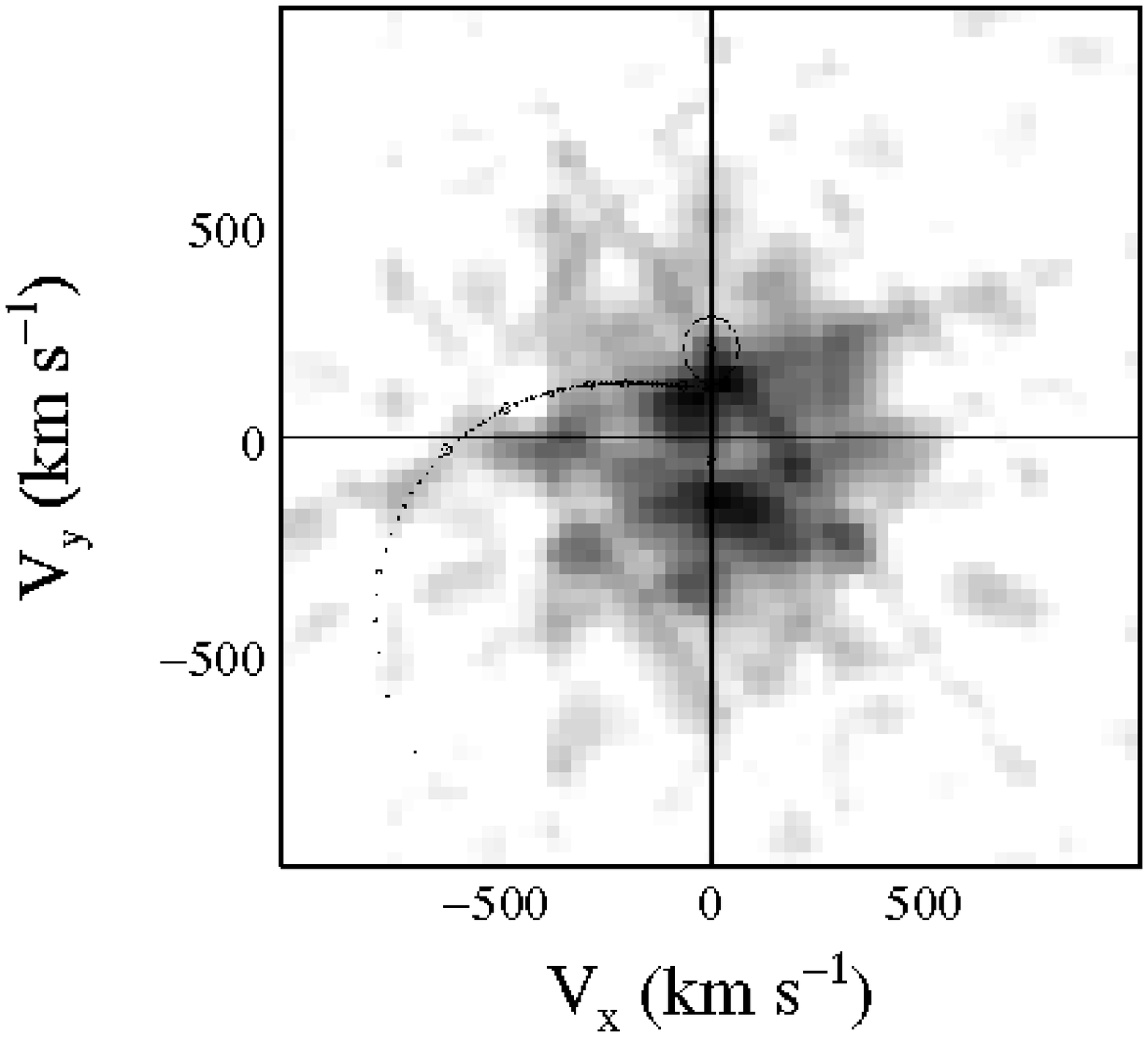,height=3.25in}\psfig{figure=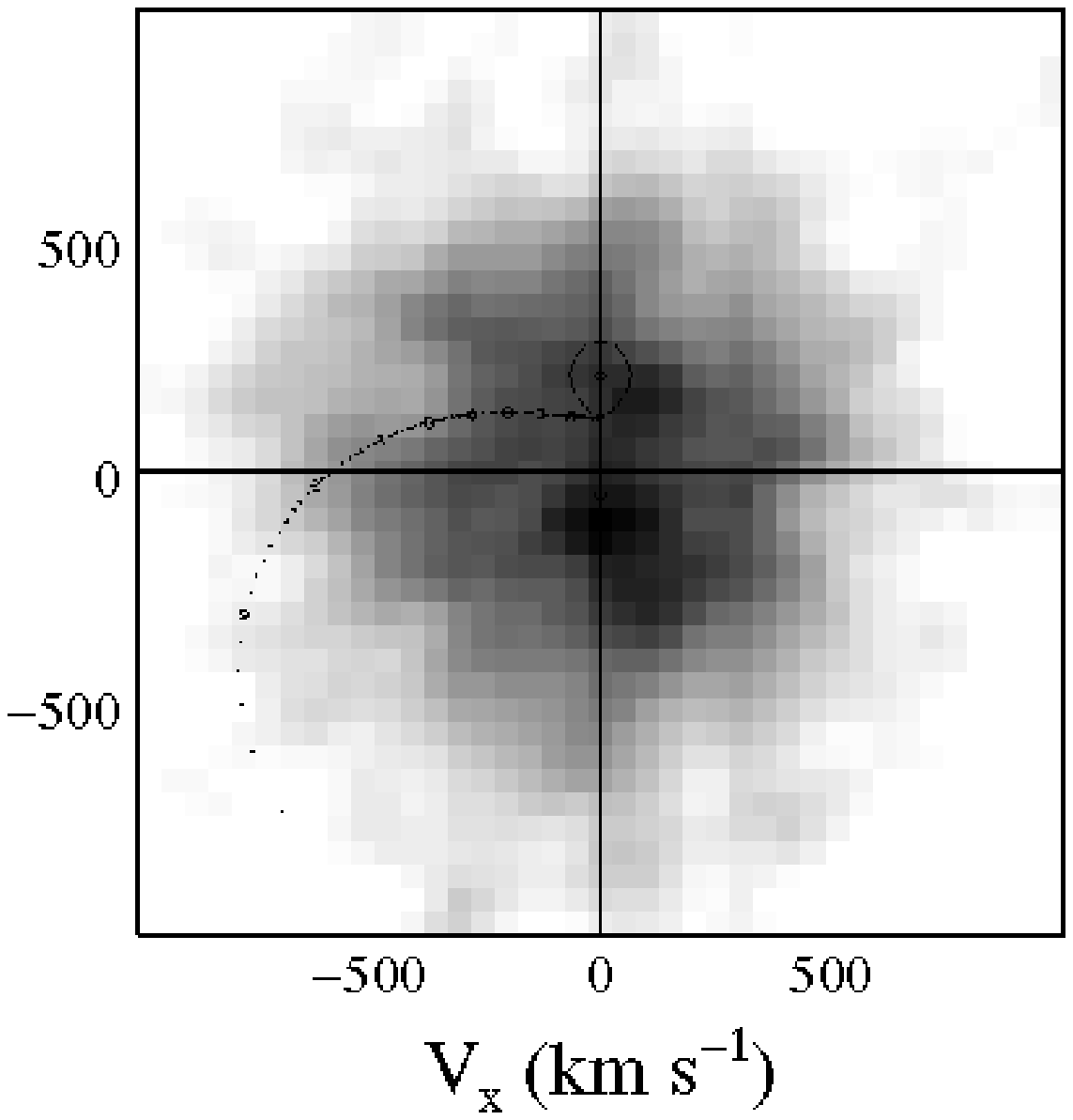,height=3.25in}}
\caption{The doppler tomograms of the H$\alpha$ (a) and H$\beta$ (b) lines
of SDSS2303 at quiescence. The oval marks the secondary, the open circle
below the origin marks the white dwarf and the dotted line marks the mass-transfer stream for a mass ratio of 0.25.}
\end{figure}

\begin{figure}
\centerline{\psfig{figure=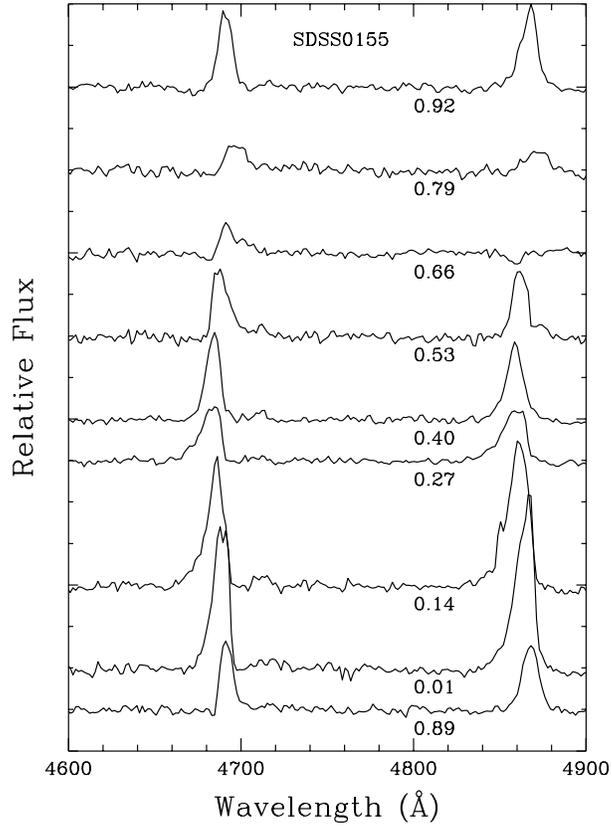,height=4.45in}}
\caption{APO spectra of the HeII 4686 and H$\beta$ lines throughout the orbital
period in the likely magnetic system SDSS0155 during a high state.}
\end{figure}

\begin{figure}
\centerline{\psfig{figure=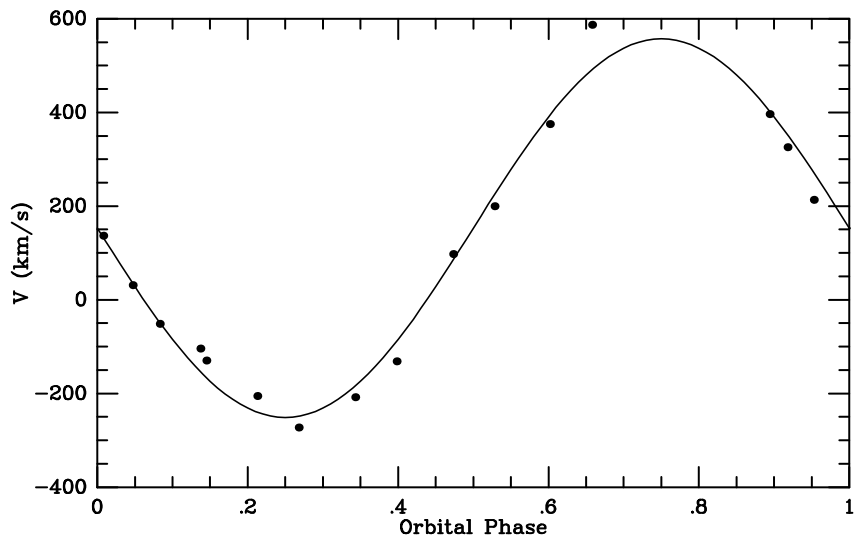,height=2.75in}}
\caption{The best fit to the HeII radial velocity curve of SDSS0155.}
\end{figure}

\begin{figure}
\centerline{\psfig{figure=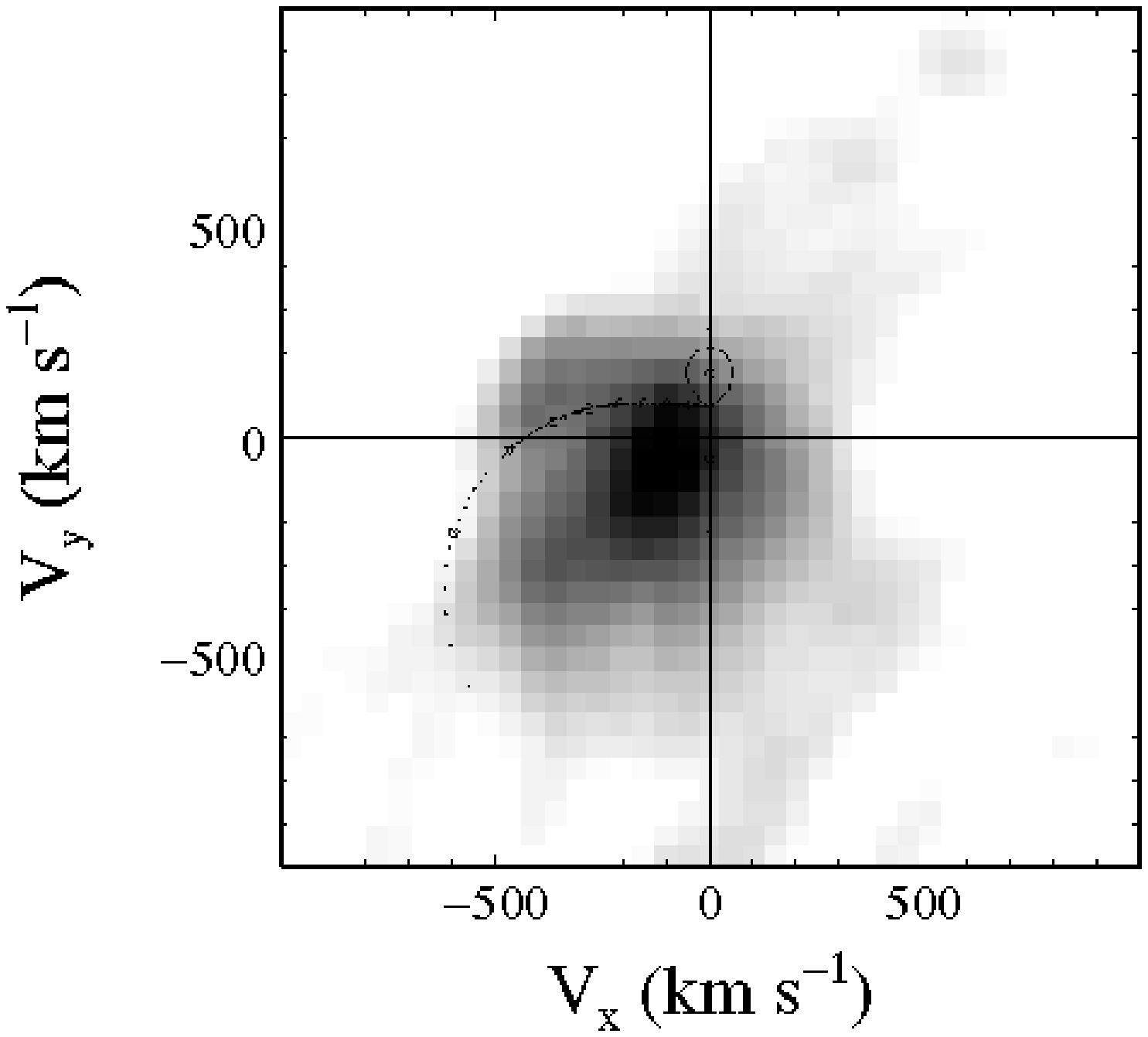,height=3.25in}\psfig{figure=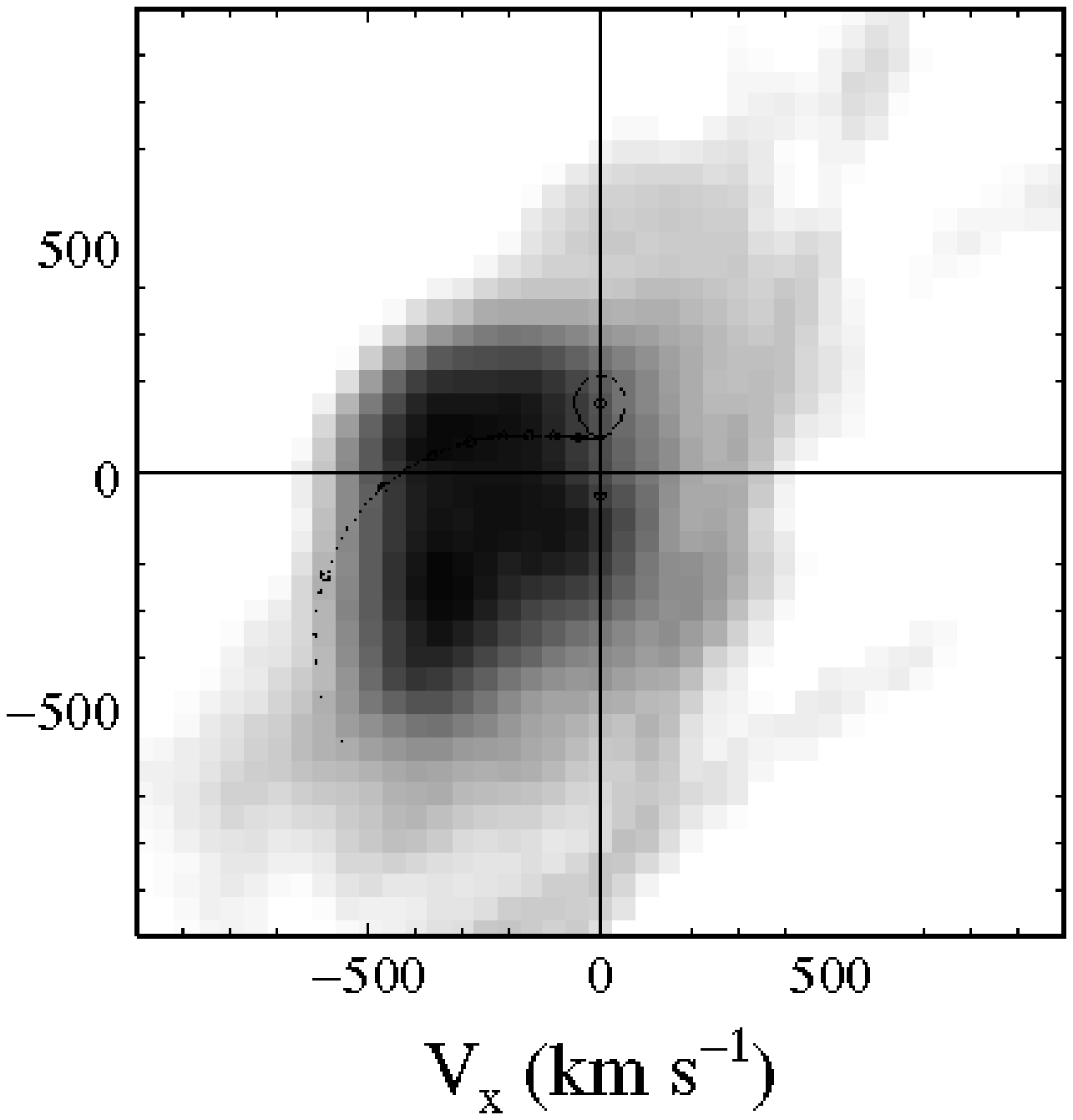,height=3.25in}}
\caption{The doppler tomograms of HeII 4686 (a) and H$\beta$ (b) for
SDSS0155. The secondary, white dwarf and mass transfer stream are shown for
a mass ratio of 0.33.}
\end{figure}

\clearpage
\scriptsize
\begin{deluxetable}{lccrrrrl}
\tablewidth{0pt}
\tablecaption{Summary of CVs with SDSS Spectra\tablenotemark{a}}
\tablehead{
\colhead{SDSSp J} &  \colhead{Date\tablenotemark{b}} & 
\colhead{$g^{*}$} & \colhead{$u^{*}-g^{*}$} & \colhead{$g^{*}-r^{*}$} & 
\colhead{$r^{*}-i^{*}$} &
\colhead{$i^{*}-z^{*}$} & \colhead{Comments\tablenotemark{c}} }
\startdata
015151.87$+$140047.2 & 11/29 & 20.26 & $-$0.29 & 0.24 & 0.42 &
0.46 & high i \nl

015543.40$+$002807.2* & 11/23 & 15.39 & 0.67 & 0.05 & $-$0.23 &
$-$0.17 & H, L, HeII, e \nl

023322.61$+$005059.5* & 10/03 & 19.94 & $-$0.30 & 0.26 & $-$0.25 &
0.01 &  \nl

072910.68$+$365838.3 & 11/29 & 20.38 & $-$0.07 & 0.34 & 0.26 &
0.05 & HeII \nl

074716.81$+$424849.0 & 12/07 & 17.10 & $-$0.02 & $-$0.09 & $-$0.09 &
$-$0.06 & NL, HeII \nl

081321.91$+$452809.4 & 11/29 & 18.29 & $-$0.05 & 0.62 & 0.46 &
0.22 & long P, K-M sec \nl

081610.84$+$453010.2 & 11/29 & 20.08 & $-$0.41 & 0.54 & 0.47 &
0.50 & M sec \nl

082236.05$+$510524.6 & 12/04 & 15.34 & 0.02 & 0.15 & $-$0.02 &
$-$0.01 & BH Lyn \nl

082409.73$+$493124.4 & 11/21 & 19.28 & $-$0.39 & 0.44 & 0.16 &
0.16 &  \nl

083642.80$+$532838.1 & 12/05 & 16.87 & 0.03 & $-$0.06 & $-$0.07 &
0.14 & SW UMa \nl

083845.23$+$491055.5 & 11/25 & 19.59 & $-$0.40 & 0.20 & $-$0.02 &
0.30 &  \nl

125641.29$-$015852.0 & 05/30 & 20.12 & 0.25 & $-$0.06 & $-$0.28 &
$-$0.15 &   \nl

143500.22$-$004606.2* & 04/03\tablenotemark{d} & 18.59 &$-$0.14 & 0.17 & $-$0.02 
& 0.10 & Vir4 \nl

155531.99$-$001055.0 & 05/28 & 19.36 & $-$0.26 & 0.23 & 0.06 &
0.31 & high i \nl

155644.24$-$000950.2 & 05/29 & 18.05 & 0.24 & 0.12 & 0.24 &
0.44 &   \nl

161033.64$-$010223.3 & 05/26 & 19.07 & 0.19 & 0.02 & $-$0.13 &
0.13 &  \nl

161332.56$-$000331.0 & 05/29 & 18.63 & $-$0.01 & 0.66 & 0.52 &
 0.31 &  \nl

163722.21$-$001957.1\tablenotemark{e} & 05/07 & 16.56 & 0.15 & $-$0.05 & $-$0.14 & 
$-$0.08 & H, L \nl

171247.71$+$604603.3*\tablenotemark{f} & 05/31\tablenotemark{g} & 19.95 & 0.92 & 0.91 & 0.59 & 0.45
&  \nl

172601.96$+$543230.7* & 09/26 & 20.52 & $-$0.26 & $-$0.04 & $-$0.02 &
0.02 &  \nl

173008.38$+$624754.7* & 05/30 & 15.92 & 0.10 & $-$0.25 & $-$0.11 &
0.00 & DN \nl

230351.64$+$010651.0 & 09/05 & 19.05 & $-$0.38 & 0.30 & 0.06 &
0.26 & DN \nl

\enddata
\tablenotetext{a}{Objects marked with asterisk are publicly available in the SDSS early data release}
\tablenotetext{b}{UT Date of spectrum (mm/dd/2000)}
\tablenotetext{c}{DN is a dwarf nova, e is eclipsing, NL is a nova-like, H,L
means shows high and low brightness states}
\tablenotetext{d}{This object also had spectra obtained on 05/26}  
\tablenotetext{e}{The correct object is the northern one of a close pair}
\tablenotetext{f}{The correct object is the southern one of a close pair}
\tablenotetext{g}{This object also had spectra obtained on 08/24}
\end{deluxetable}

\begin{deluxetable}{llrrrrrrrrrr}
\tablewidth{0pt}
\tablecaption{SDSS Spectral Line Fluxes and Equivalent Widths\tablenotemark{a}}
\tablehead{
\colhead{SDSS} & \colhead{Plate-Fiber} & \multicolumn{2}{c}{H$\gamma$} & 
\multicolumn{2}{c}{H$\beta$} &
\multicolumn{2}{c}{H$\alpha$} &
\multicolumn{2}{c}{He4471} & \multicolumn{2}{c}{HeII4686}\\
\colhead{} & \colhead{} & \colhead{F} & \colhead{EW} & \colhead{F} &
\colhead{EW} & \colhead{F} & \colhead{EW} & \colhead{F} & \colhead{EW} &
\colhead{F} & \colhead{EW} } 
\startdata
0151 & 430-430 & 0.9 & 15 & 1.7 & 36 & 2.6 & 77 & & &  & \nl
0155 & 403-423\tablenotemark{b} & 59.7 & 10 & 63.4 & 11 & 37.8 & 13 & 12.2 & 2
 & 44.5 & 8 \nl
0233 & 407-592 & 1.9 & 43 & 2.5 & 74 & 4.5 & 303 & 0.2 & 5 & &  \nl
0729 & 431-316 & 0.5 & 17 & 0.6 & 24 & 0.8 & 35 & 0.07 & 3 & 0.5 & 19 \nl
0747 & 434-430 & 0.4 & 0.8 & 0.6 & 1.2 & 0.9 & 3.3 & & & 0.8 & 1.5 \nl
0813 & 439-582 & 1.6 & 15 & 1.9 & 14 & 2.8 & 15 & & & & \nl
0816 & 439-624 & 3.5 & 43 & 3.8 & 53 & 5.3 & 91 & 0.9 & 12 & & \nl
0824 & 443-360 & 1.0 & 17 & 1.5 & 26 & 1.9 & 46 & & & & \nl
0838 & 445-89 & 2.9  & 65 & 3.2 & 87 & 5.1 & 211 & 0.4 & 8 & & \nl
1256 & 338-382 & 0.4 & 11 & 0.4 & 14 & 0.7 & 52 & & & & \nl
1435\tablenotemark{c} & 306-4 & 2.9 & 19 & 2.5 & 23 & 4.4 & 64 &  & & & \nl
1435\tablenotemark{d} & 306-20 & 22.4 & 59 & 18.3 & 66 & 19.0 & 105 & 3.9 & 11 
& &  \nl  
1555 & 343-198 & 4.9 & 33 & 5.9 & 62 & 10.1 & 195 & 0.8 & 5.5 & &  \nl
1556 & 344-315 & 1.7 & 6 & 2.8 & 14 & 6.2 & 62 & & & & \nl 
1610 & 345-138 & 0.9 & 5 & 1.9 & 16 & 5.1 & 86 & & & & \nl
1613 & 346-263 & 5.1 & 11 & 5.4 & 14 & 8.3 & 27 & & & & \nl
1637 & 348-103 & 2.6 & 80 & 3.0 & 129 & 3.4 & 186 & 0.6 & 21 & &  \nl
1712 & 351-17 & 6.0 & 100 & 6.8 & 79 & 8.3 & 90 & 2.0 & 28 & 2.0 & 23 \nl 
1726 & 357-51 & 1.4 & 56 & 1.6 & 94 & 1.3 & 158 & 0.3 & 13 & 0.25 & 12 \nl
1730 & 352-26 & 16.1 & 58 & 15.4 & 63 & 17.5 & 118 & 3.4 & 14 & 1.3 & 6 \nl
2303 & 380-575 & 2.8 & 52 & 3.3 & 74 & 5.1 & 193 & 0.5 & 10 & &  \nl
\enddata
\tablenotetext{a}{Fluxes are in units of 10$^{-15}$ ergs cm$^{-2}$ s$^{-1}$,
equivalent widths are in units of \AA}
\tablenotetext{b}{Measurements from APO spectra as SDSS spectrum saturated}
\tablenotetext{c}{Measured from spectrum on 04/03}
\tablenotetext{d}{Measured from spectrum on 05/26}
\end{deluxetable}

\normalsize
\begin{deluxetable}{lcccl}
\tablewidth{0pt}
\tablecaption{Followup Data from MRO and APO}
\tablehead{
\colhead{UT Date} & \colhead{SDSS} & \colhead{Site} &
\colhead{Interval} & \colhead{Data Obtained} }
\startdata
05/06/00 & 1435 & APO & 0.3 hrs & 2 spectra at quiescence \nl
05/15/00 & 1435 & APO & 10 min & 1 spectrum at quiescence \nl

05/28/00 & 1637 & APO & 10 min & 1 spectrum at quiescence \nl
07/03/00 & 1637 & APO & 15 min & 1 spectrum at quiescence \nl

06/05/00 & 1730 & MRO & 5 hrs & V photometry \nl
06/16/00 & 1730 & MRO & 1.75 hrs & V photometry \nl
07/16/00 & 1730 & MRO & 2.4 hrs & V photometry \nl
08/07/00 & 1730 & MRO & 4.8 hrs & V photometry \nl
08/08/00 & 1730 & MRO & 7.2 hrs & B photometry \nl
08/25/00 & 1730 & MRO & 2 hrs & B photometry \nl
10/03/00 & 1730 & APO & 2.2 hrs & 20 spectra at outburst \nl
10/06/00 & 1730 & APO & 0.25 hr & 2 spectra at quiescence \nl

10/03/00 & 2303 & APO & 1.5 hrs & 9 spectra at quiescence \nl
10/06/00 & 2303 & APO & 1.9 hrs & 10 spectra at outburst \nl

12/29/00 & 0155 & APO & 3 hrs & 17 spectra at high state \nl

01/15/01 & 0747 & APO & 10 min & 1 spectrum at high state \nl 
\enddata
\end{deluxetable}

\begin{deluxetable}{lcccccc}
\tablewidth{0pt}
\tablecaption{Radial Velocity Solutions}
\tablehead{
\colhead{SDSS} & \colhead{Line} & \colhead{P (min)} & 
\colhead{$\gamma$} & \colhead{K (km/s)} &
\colhead{$\phi$} & \colhead{$\sigma$} }
\startdata
1730 & H$\alpha$ & 117$\pm$5 & 23$\pm$3 & 74$\pm$6 & 0.0$\pm$0.02 & 19 \nl
2303 & H$\alpha$ & 100$\pm$14 & 9.3$\pm$0.1  & 53$\pm$7 & 0.0$\pm$0.02 
& 14 \nl
2303 & H$\beta$ & 100 & 55.4$\pm$0.1 & 49$\pm$5  & 0.11$\pm$0.02 & 9 \nl
0155 & H$\alpha$ & 86$\pm$2 & 89$\pm$8 & 359$\pm$29 & 0.09$\pm$0.01 & 64 \nl
0155 & H$\beta$ & 87$\pm$2 & 135$\pm$5 & 406$\pm$20 & 0.05$\pm$0.01 & 45 \nl
0155 & HeII4686 & 88$\pm$2 & 153$\pm$4 & 404$\pm$18 & 0.00$\pm$0.01 & 38 \nl
\enddata
\end{deluxetable}

\scriptsize
\begin{deluxetable}{lccc}
\tablewidth{0pt}
\tablecaption{ROSAT Detections}
\tablehead{
\colhead{SDSS} & \colhead{ROSAT (c/s)\tablenotemark{a}} & \colhead{Exp (sec)}
& \colhead{Obs\tablenotemark{b}} }
\startdata
0155 & 0.04$\pm$0.01 & 394 & S \nl
0233 & 0.0058$\pm$0.0005 & 28220 & P \nl
0838 & 0.017$\pm$0.003 & 4569 & P+S \nl
1730 & 0.060$\pm$0.006 &  2644 & S \nl
2303 & 0.061$\pm$0.015 & 348 & S \nl
\enddata
\tablenotetext{a}{For a 2 keV bremsstrahlung spectrum, 1 c/s corresponds to a
0.1-2.4 keV flux of about 7$\times10^{-12}$ ergs cm$^{-2}$ s$^{-1}$}
\tablenotetext{b}{S signifies a survey observation; P 
a pointed observation}
\end{deluxetable}

\begin{deluxetable}{lccc}
\tablewidth{0pt}
\tablecaption{2MASS Detections}
\tablehead{
\colhead{SDSS} & \colhead{J} & \colhead{H} & \colhead{K} } 
\startdata
0747 & 16.59$\pm$0.13 & 16.31$\pm$0.22 & 15.5\tablenotemark{a} \nl
0813 & 15.99$\pm$0.09 & 15.30$\pm$0.09 & 15.19$\pm$0.13 \nl
1712 & 16.80$\pm$0.16 & 15.94$\pm$0.19 & 15.5\tablenotemark{a} \nl
1730 & 15.30$\pm$0.05 & 15.20$\pm$0.09 & 15.28$\pm$0.19 \nl
\enddata
\tablenotetext{a}{Detection only}
\end{deluxetable}

\begin{deluxetable}{lrrrc}
\tablewidth{0pt}
\tablecaption{LONEOS Observations}
\tablehead{
\colhead{SDSS} & \colhead{No.Obs} & \colhead{Days} & 
\colhead{Mag Range\tablenotemark{a}} & \colhead{Error} }
\startdata
0155 & 21 & 8 & 14.7 - 17.6 & 0.05 \nl
0747 & 6 & 4 & 17.04 - 17.10 & 0.15\nl
0824 & 7 & 2 & 17.60 - 17.75 & 0.1\nl
2303 & 14 & 6 & 17.3 - 18.6 & 0.1 \nl
\enddata
\tablenotetext{a}{Calibration from USNO-A2.0 red magnitudes} 
\end{deluxetable}

\end{document}